# Significantly Reduced Thermal Conductivity in $\beta$-(Al$_{0.1}$Ga$_{0.9}$)$_2$O$_3$/Ga$_2$O$_3$ Superlattices

Zhe Cheng,[1] Nicholas Tanen,[2] Celesta Chang,[3] Jingjing Shi,[1] Jonathan McCandless,[4] David Muller,[5,6] Debdeep Jena,[2,4,6] Huili Grace Xing,[2,4,6] Samuel Graham,[1,7, a)]

[1] George W. Woodruff School of Mechanical Engineering, Georgia Institute of Technology, Atlanta, Georgia 30332, USA

[2] Department of Materials Science and Engineering, Cornell University, Ithaca, New York 14853, USA

[3] Department of Physics, Cornell University, Ithaca, New York 14853, USA

[4] School of Electrical and Computer Engineering, Cornell University, Ithaca, New York 14853, USA

[5] School of Applied and Engineering Physics, Cornell University, Ithaca, New York 14853, USA

[6] Kavli Institute for Nanoscale Science, Cornell University, Ithaca, New York 14853, USA

[7] School of Materials Science and Engineering, Georgia Institute of Technology, Atlanta, Georgia 30332, USA

a) Corresponding author: sgraham@gatech.edu

Abstract

β-Ga$_2$O$_3$ has emerged as a promising candidate for electronic device applications because of its ultra-wide bandgap, high breakdown electric field, and large-area affordable substrates grown from the melt. However, its thermal conductivity is at least one order of magnitude lower than that of other wide bandgap semiconductors such as SiC and GaN. Thermal dissipation in electronics made from β-Ga$_2$O$_3$ will be the bottleneck for real-world applications, especially for high power and high frequency devices. Similar to GaN/AlGaN interfaces, $β$-(Al$_x$Ga$_{1-x}$)$_2$O$_3$/Ga$_2$O$_3$ heterogeneous structures have been used to form a high mobility two-dimensional electron gas (2DEG) where joule heating is localized. The thermal properties of $β$-(Al$_x$Ga$_{1-x}$)$_2$O$_3$/Ga$_2$O$_3$ are the key for heat dissipation in these devices while they have not been studied before. This work reports the first measurement on thermal conductivity of $β$-(Al$_{0.1}$Ga$_{0.9}$)$_2$O$_3$/Ga$_2$O$_3$ superlattices from 80 K to 480 K. Its thermal conductivity is significantly reduced (5.7 times reduction) at room temperature comparing with that of bulk Ga$_2$O$_3$. Additionally, the thermal conductivity of bulk Ga$_2$O$_3$ with (010) orientation is measured and found to be consistent with literature values regardless of Sn doping. We discuss the phonon scattering mechanism in these structures by calculating their inverse thermal diffusivity. By comparing the estimated thermal boundary conductance (TBC) of $β$-(Al$_{0.1}$Ga$_{0.9}$)$_2$O$_3$/Ga$_2$O$_3$ interfaces and Ga$_2$O$_3$ maximum TBC, we reveal that some phonons in the superlattices transmit through several interfaces before scattering with other phonons or structural imperfections. This study is not only important for Ga$_2$O$_3$ electronics applications especially for high power and high frequency applications, but also for the fundamental thermal science of phonon transport across interfaces and in superlattices.

Introduction

As an emerging ultra-wide bandgap semiconductor material, β-Ga$_2$O$_3$ has shown promising properties for electronic device applications, such as an ultra-wide bandgap (4.8 eV) and high critical electric field (8 MV/cm), which predicts a Baliga figure of merit that is 3214 times that of Si.[1] However, the thermal conductivity of bulk β-Ga$_2$O$_3$ (10-30 W/m-K, depending on crystal orientation) is at least one order of magnitude lower than those of other wide bandgap semiconductors such as GaN (230 W/m-K), 4H-SiC (490 W/m-K), and diamond (>2000 W/m-K).[2,3] Thermal dissipation will be the bottleneck for real-world applications, especially for high power and high frequency devices. Currently, compared to demonstrations of Ga$_2$O$_3$ devices, a disproportionately smaller number of thermal studies have been performed.[4] Similar to GaN/AlGaN interfaces, to demonstrate modulation-doped field effect transistors (MODFETs), β-(Al$_x$Ga$_{1-x}$)$_2$O$_3$/Ga$_2$O$_3$ heterogeneous structures have been used to form a high mobility two-dimensional electron gas (2DEG) where joule heating is localized.[5-9] The thermal properties of the β-(Al$_x$Ga$_{1-x}$)$_2$O$_3$/Ga$_2$O$_3$ structure are the key for heat dissipation in these devices, however they have not been studied before.

In this work, we report the first measurement on temperature-dependent thermal conductivity of β-(Al$_{0.1}$Ga$_{0.9}$)$_2$O$_3$/Ga$_2$O$_3$ superlattices epitaxial-grown on bulk (010) Ga$_2$O$_3$ substrates by molecular-beam epitaxy (MBE) from 80 K to 480 K. Multi-frequency time-domain thermoreflectance (TDTR) is used to measure the thermal properties of both the β-(Al$_{0.1}$Ga$_{0.9}$)$_2$O$_3$/Ga$_2$O$_3$ superlattices and the bulk Ga$_2$O$_3$ substrates simultaneously. The phonon scattering mechanism in these structures is discussed in detail. Additionally, we estimate the

TBC of $\beta$-(Al$_{0.1}$Ga$_{0.9}$)$_2$O$_3$/Ga$_2$O$_3$ interfaces and compare it with maximum Ga$_2$O$_3$ TBC. The mechanism of phonons transmission through interfaces is discussed.

Methods

The $\beta$-(Al$_x$Ga$_{1-x}$)$_2$O$_3$/Ga$_2$O$_3$ superlattice used in this study was homoepitaxy-grown on a Sn-doped (010) Ga$_2$O$_3$ substrate, with an n-type doping concentration of $4\times10^{18}$ cm$^{-3}$, using a Veeco Gen930 MBE system. The aluminum and gallium were provided by standard effusion cells. The oxygen plasma was produced using a Veeco RF plasma source. During the growth, the gallium and aluminum beam equivalent pressures (BEP) measured by an ion gauge were $1 \times 10^{-8}$ Torr and $1 \times 10^{-9}$ Torr, respectively. This led to an aluminum flux that is 9.1% of the total metal flux. The oxygen was flown into the chamber at 0.7 sccm and the RF plasma was struck using a load power of 289 W, which corresponded to a total chamber pressure of $2.18 \times 10^{-5}$ Torr. The substrate temperature, measured by a thermocouple, was 500 °C for the entire growth. The substrate was mounted to a silicon carrier wafer using indium bonding. The substrate was grown by edge-defined film-fed growth (EFG) purchased from Novel Crystal Technology. The film has ten alternating periods of a Ga$_2$O$_3$ layer followed by a (Al$_{0.1}$Ga$_{0.9}$)$_2$O$_3$ layer each grown for 30 minutes.

Cross-sectional TEM specimen was prepared using an FEI Strata 400 Focused Ion Beam (FIB) with a final milling step of 5keV to reduce surface damage. Atomic resolution high angle annular dark field (HAADF) images were acquired on an aberration corrected 300keV Themis Titan. The superlattice film thickness was determined by a scanning transmission electron microscopy (STEM) to be 114 nm, as shown in Figure 1. For each period, the Ga$_2$O$_3$ layer is about 6.5 nm thick while the (Al$_{0.1}$Ga$_{0.9}$)$_2$O$_3$ layer is about 4.5 nm thick. The interfaces in the superlattice structure are not very sharp. A layer of Al (~80 nm) was deposited on the surface as the TDTR transducer as demonstrated in several previous studies.[10-14] A modulated pump beam heats the sample surface periodically while a delayed probe beam detects the surface temperature variation via thermoreflectance. The measured signal of temperature variation is fit with an analytical heat transfer solution to infer unknown parameters. Here, we measure one spot on the sample with two modulation frequencies, i.e., 3.6 MHz and 8.8 MHz, at room temperature. TDTR signal is sensitive to the thermal conductivity of the bulk $Ga_2O_3$ substrate with modulation frequency of 3.6 MHz while it is sensitive to the thermal conductivity of the superlattice with modulation frequency of 8.8 MHz. More details about multi-frequency TDTR measurements can be found in references.[10,15-17] Additionally, we estimate the error bars here as ±10%.

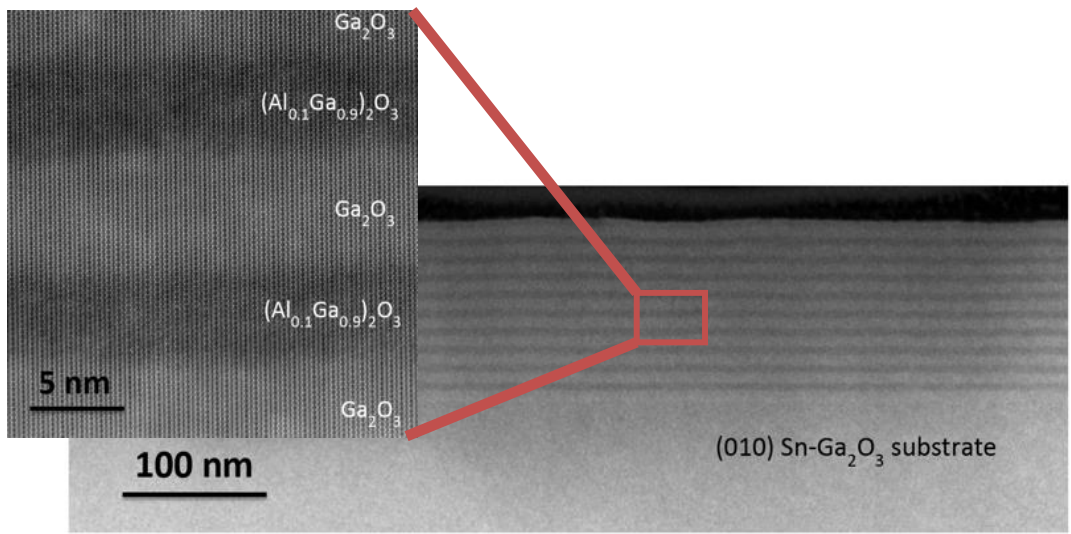

Figure 1. HAADF-STEM images of the $\beta$ -$(Al_{0.1}Ga_{0.9})_2O_3$/$Ga_2O_3$ superlattice structure.

Results and Discussion

The temperature-dependent thermal conductivity of the bulk (010) $Ga_2O_3$ substrate and the $\beta$-$(Al_{0.1}Ga_{0.9})_2O_3$/$Ga_2O_3$ superlattice are shown in Figure 2 (a). At room temperature, the thermal conductivity of the superlattice is 5.7 times smaller than that of the bulk (010) $Ga_2O_3$ substrate. This significantly reduced thermal conductivity further impedes thermal dissipation, potentially creating additional challenges for gallium oxide electronics devices. Aggressive thermal management techniques need to be applied for reliable device performance, such as integrating high thermal conductivity materials close to the regions where heat is being generated to aid in pulling out the heat. For the bulk (010) $Ga_2O_3$ substrate, its thermal conductivity decreases with increasing temperature from 80 K to 450 K because of increased phonon-phonon scattering. As temperature increases, the number of excited phonons increases, resulting in increasingly extensive phonon-phonon scatterings. For the superlattice, its thermal conductivity shows a peak at 380 K. Below 380 K, the thermal conductivity decreases with decreasing temperature while it decrease with increasing temperature above 380 K. For temperatures below 380 K, phonon-structural imperfection scattering, such as alloy and boundary, dominates in impeding thermal transport. For temperatures above 380 K, phonon-phonon scattering dominates. More about the scattering mechanisms will be discussed later.

The measured thermal conductivity of the bulk (010) $Ga_2O_3$ substrate is compared with literature values as shown in Figure 2 (b). Our measured values are consistent with most of other experimentally measured values and first-principle calculated values in the literature. This indicates that the Sn doping and unintentional doping do not affect the thermal conductivity significantly. The other samples measured in the literature have different levels of doping as well. It is still an open question as to how high doping concentrations would affect the thermal conductivity in β-$Ga_2O_3$.

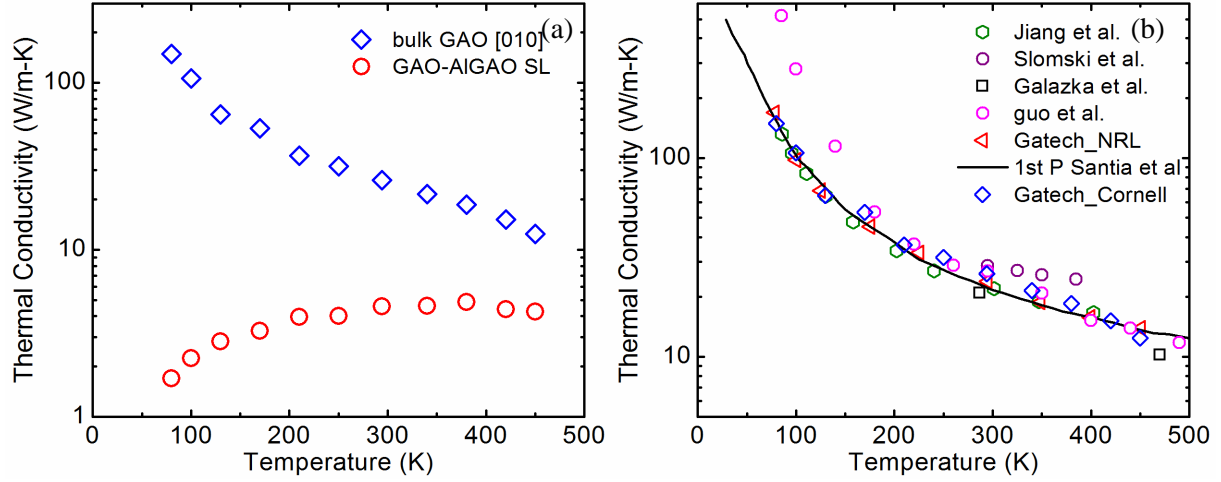

Figure 2. (a) Temperature-dependent thermal conductivity of the bulk $Ga_2O_3$ substrate and $\beta$-$(Al_{0.1}Ga_{0.9})_2O_3/Ga_2O_3$ superlattices. (b) Summary of temperature-dependent thermal conductivity of bulk (010) $\beta$-$Ga_2O_3$ in this work and literature.[18-20]

To better understand the phonon scattering mechanism in the superlattice and bulk $Ga_2O_3$, we calculate the inverse thermal diffusivity of both the superlattice and bulk $Ga_2O_3$, as shown in Figure 3 (a). The formula of inverse thermal diffusivity is shown as below:

$$\frac{C_v}{k}(T) = \left\{ \sum_{n=1}^{n} \int_0^{\omega_D} g(\omega) k_B \left(\frac{\hbar\omega}{k_B T}\right)^2 e^{\frac{\hbar\omega}{k_B T}} / (e^{\frac{\hbar\omega}{k_B T}} - 1)^2 \, d\omega \right\} /$$

$$\left\{ \sum_{n=1}^{n} \int_0^{\omega_D} g(\omega) k_B \left(\frac{\hbar\omega}{k_B T}\right)^2 e^{\frac{\hbar\omega}{k_B T}} / (e^{\frac{\hbar\omega}{k_B T}} - 1)^2 v_\omega^2 \left(\frac{1}{\tau_{ph}} + \frac{1}{\tau_{struc}}\right)^{-1} d\omega \right\}. \quad (1)$$

Here, $C_v$ is the volumetric heat capacity, $k$ is thermal conductivity, $T$ is temperature, $n$ is the number of phonon branches, $\omega_D$ is Debye frequency, $g(\omega)$ is phonon density of states, $k_B$ is Boltzmann constant, $\hbar$ is reduced Planck constant, $\omega$ is the phonon frequency, $v_\omega$ is the phonon group velocity, $\tau_{ph}$ is the relaxation time for phonon-phonon scattering, and $\tau_{struc}$ is the relaxation time for phonon-structural imperfection scattering. To the first-order approximation, the inverse thermal diffusivity can be used to estimate the relative contribution of scattering sources.[21,22] The temperature-dependent thermal conductivity of the superlattice and bulk $Ga_2O_3$ have different trends with temperature because of the strong temperature dependence of heat capacity. As a result, it is difficult to compare the contribution of phonon-phonon scattering and phonon-structural imperfection scattering according to the thermal conductivity data. After removing the effect of heat capacity, the inverse thermal diffusivity represents the relative contributions of phonon scattering sources qualitatively. As shown in Figure 3 (a), the inverse thermal diffusivity decreases with decreasing temperature because of reduced scattering intensity of phonon-phonon scattering. As temperature goes to zero, phonon-phonon scatterings diminish and only structural imperfection scatterings remain. Then Equation (1) can be simplified as:

$$\frac{C_v}{k}(T \to 0 \text{ K}) \approx \frac{3}{v_0^2 \tau_{struc}} = \frac{3}{v_0 l_0}. \quad (2)$$

Here, $v_0$ is the average phonon group velocity and $l_0$ is the scattering length arising from structural imperfections. For the bulk $Ga_2O_3$, the structural imperfection is negligible so nearly zero residual value is observed at low temperatures. However, a large residual value is observed at low temperatures for the superlattice, indicating strong structural imperfection scatterings, such as alloy scattering, boundary scattering, and point defect scattering. We estimate the average acoustic phonon group velocity as 2420 m/s.[20] Based on the residual inverse thermal diffusivity (40 s/cm²), the structural scattering length $l_0$ can be calculated as 3.1 nm, which is close to the layer thickness of the superlattice by considering additional alloy scatterings. The rough interfaces in the superlattice may reduce the effective thermal conductivity because the roughness at the interfaces may induce additional phonon scatterings.[23]

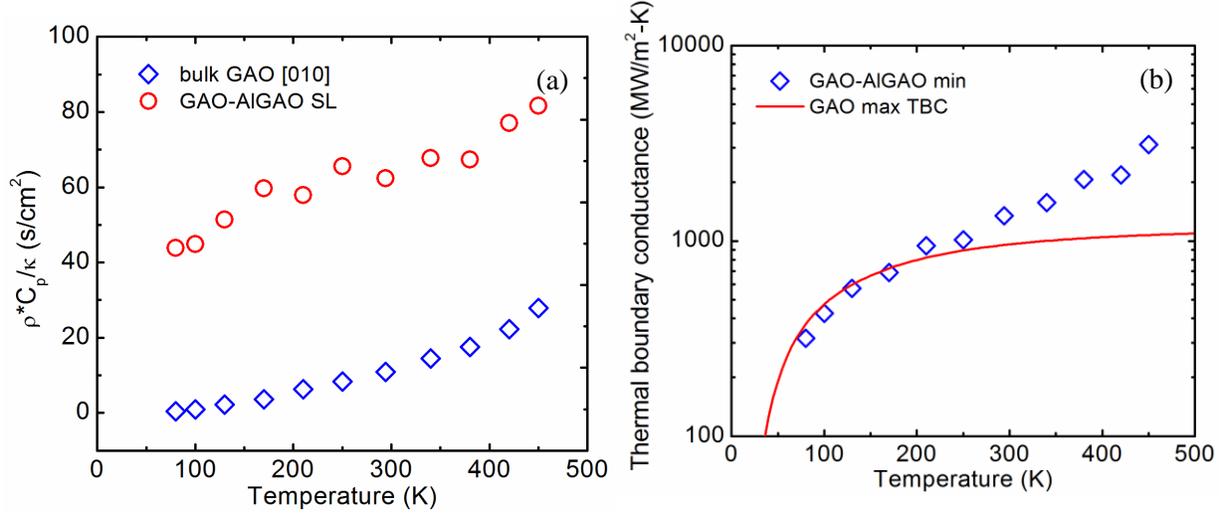

Figure 3. (a) Temperature-dependent inverse thermal diffusivity of bulk (010) $Ga_2O_3$ and $\beta$-$(Al_{0.1}Ga_{0.9})_2O_3/Ga_2O_3$ superlattices. (b) Temperature dependence of estimated minimum TBC (lower bound) of $\beta$-$(Al_{0.1}Ga_{0.9})_2O_3$ and $Ga_2O_3$ interfaces and maximum TBC of $Ga_2O_3$.

To understand the mechanism of phonons transmission through $(Al_{0.1}Ga_{0.9})_2O_3/Ga_2O_3$ interfaces, we estimate the TBC as below:

$$\frac{d_1}{k_{GAO}} + \frac{d_2}{k_{AlGAO}} + \frac{2}{TBC} = \frac{d_1+d_2}{k_{measured}}. \qquad (3)$$

Here, $d_1$ and $d_2$ are the thicknesses of the $Ga_2O_3$ layer and the $(Al_{0.1}Ga_{0.9})_2O_3$ layer in one period. $k_{GAO}$ and $k_{AlGAO}$ are the thermal conductivity of the $Ga_2O_3$ layer and the $(Al_{0.1}Ga_{0.9})_2O_3$ layer. TBC in this case refers to the TBC of the $(Al_{0.1}Ga_{0.9})_2O_3/Ga_2O_3$ interface. $k_{measured}$ is the measured effective thermal conductivity of the superlattice. $k_{GAO}$ and $k_{AlGAO}$ are supposed to be much lower than the bulk thermal conductivity because of boundary scattering and alloy scatterings. Here, we only consider the size effect resulting from total superlattice thickness (114 nm), the thermal conductivity reduces to 45% according to first-principle calculations.[20] As a result, the minimum TBC of the $(Al_{0.1}Ga_{0.9})_2O_3/Ga_2O_3$ interface could be estimated by assuming $k_{GAO}$ and $k_{AlGAO}$ as bulk values only with size effect from the total thickness. To compare with this estimated TBC, we also calculate the max TBC of any heterogeneous interfaces which involving $Ga_2O_3$. This max TBC is calculated by assuming the phonon transmission coefficient across the interface as unity (all phonons from $Ga_2O_3$ could transmit through the interface). As shown in Figure 3 (b), the estimated minimum TBC is larger than the max TBC of $Ga_2O_3$ interfaces, especially at high temperatures. At 450 K, the minimum TBC is almost three times larger than the max $Ga_2O_3$ TBC. This means that some phonons transmit through several interfaces before scattering with other phonons or structural imperfections. Here, the estimated TBC is the thermal energy transmitted across the interface for a certain temperature difference per unit area. Because the period of the superlattice is very small comparing with some long mean free path phonons, some phonons could transmit several interfaces without scattering with other phonons and structural imperfections. The energy of these phonons are accounted repeatedly for several times, resulting in a very large effective TBC, even larger than the max TBC $Ga_2O_3$ hetero-interfaces could achieve. This non-local and non-equilibrium phonon transport across interfaces is one of the challenges to define local temperature and understand thermal transport across interfaces. Similar phenomenon was observed in AlN-GaN superlattices before.[24] Our work is important since it may be necessary to design the superlattices not only for the creation of the channel 2DEG, but also for more efficient phonon dissipation through the structure. This electro-thermal co-design is truly an important feature for future wide bandgap devices which require enhancements in heat dissipation within the devices.

## Conclusions

This work reports the first temperature-dependent measurement on thermal conductivity of $\beta$-$(Al_{0.1}Ga_{0.9})_2O_3/Ga_2O_3$ superlattices from 80 K to 480 K. We observed significantly reduced thermal conductivity (5.7 times reduction) at room temperature comparing with bulk $Ga_2O_3$. The thermal conductivity of bulk (010) $Ga_2O_3$ is measured and found to be consistent with literature values. By calculating the inverse thermal diffusivity of both the superlattice and bulk $Ga_2O_3$, we qualitatively identify the relative contribution of scattering intensity of phonon-phonon scattering and phonon-structural imperfection scattering. We estimated the scattering length as 3.1 nm, which is close to the layer thickness of the superlattice by considering additional alloy scatterings. The estimated minimum TBC of $\beta$-$(Al_{0.1}Ga_{0.9})_2O_3/Ga_2O_3$ interfaces is found to be larger than the $Ga_2O_3$ maximum TBC. This result shows that some phonons could transmit through several interfaces before scattering with other phonons or structural imperfections. This study is not only important for $Ga_2O_3$ electronics applications especially for high power and high frequency applications, but also for the fundamental thermal science of phonon transport across interfaces and in superlattices.

## Acknowledgement

The authors would like to acknowledge the funding support from Air Force Office of Scientific Research under a MURI program (Grant No. FA9550-18-1-0479) and a Center of Excellence program (Grant No. FA9550-18-1-0529) and the Office of Naval Research under a MURI program (Grant No. N00014-18-1-2429).